\documentclass[onecolumn]{aastex631}

\usepackage{bm}
\usepackage{amstext}
\usepackage{epstopdf}

\defcitealias{solomon:87}{S87}

\graphicspath{{./}{figures/}}

\begin{document}

\title{Rotation of polarization angle in gamma-ray burst prompt phase}

\correspondingauthor{}
\email{lanmixiang@jlu.edu.cn}

\author{Hao-Bing Wang}
\author[0000-0001-5641-2598]{Mi-Xiang Lan}
\affil{Center for Theoretical Physics and College of Physics, Jilin University, Changchun, 130012, China; lanmixiang@jlu.edu.cn \\}

\begin{abstract}

The rotations of the polarization angle (PA) with time (energy) can lead to the depolarization of the time-integrated (energy-integrated) polarization. However, we don't know how and when it will rotate. Here, we consider the magnetic reconnection model to investigate the polarizations, especially the PA rotations of GRB prompt emission. For the large-scale ordered aligned magnetic field configuration, we find that PAs will evolve with time (energy) for off-axis observations. Our studies show that the rotations of the PAs are due to the changes of the ``observed shape'' of the emitting region (before averaged). We apply our models to the single pulse burst of GRB 170101A and GRB 170114A with time-resolved PA observations. We find it can interpret the violent PA variation of GRB 170101A. The model could not predict the twice $90^{\circ}$ PA changes in GRB 170114A. Detailed model should be considered.


\end{abstract}

\section{Introduction}\label{intro}

Gamma-ray bursts (GRBs) are the bursts of high-energy electromagnetic radiation in the universe. GRBs can be divided into long and short bursts with a duration seperation of two seconds. Long bursts originate from the collapse of the core of a massive star (Mazzali et al. 2013; Woosley 1993; Bloom et al. 1999; MacFadyen et al. 2001; Hjorth et al. 2003). Short bursts result from the merger of two neutron stars (NSs) or a NS and a black hole (BH) (Narayan et al. 1992; Abbott et al. 2017; Goldstein et al. 2017; Lazzati et al. 2018). Despite more than 20 years of research (since 1997), the emission mechanism of GRBs remains unknown. In the prompt phase of a GRB, the observed spectral shape is a broken power law, usually described by the empirical formula ``Band function'', which has two power laws, smoothly connected at the photon energy $E_{pk}$, which is the peak of $\nu F_{\nu}$ (Band et al. 1993).

To explain the origin of the prompt emission from a GRB, the internal shock model was considered (Paczynski \& Xu 1994; Ress \& Meszaros 1994; Kobayashi et al. 1997; Sari \& Piran 1997; Daigne \& Mochkovitch 1998). Another popular model of GRB prompt emission is the magnetic reconnection model (zhang \& Yan 2011), this model assumes that the central engine is highly magnetized, and the ejecting shells are also highly magnetized. The collisions of these shells distort the magnetic field, resulting in a magnetic reconnection event, in which electrons in the reconnected region are accelerated to produce synchrotron radiation. Although the acceleration processes of the electrons in two models are different, both can explain the typical spectra of the GRBs.

Many works have given results on time-integrated and energy-integrated polarization (e.g., Toma et al. (2009); Guan \& Lan (2022)),however, informations on the evolution of polarization are missing. The evolution of polarization is important to constrain the physical process of GRB prompt emission. Studies on time-resolved and energy-resolved polarization have been done until recently (Lan \& Dai 2020; Lan et al. 2021). But these studies do not predict the rotations of the polarization angle (PA).

\cite{Zhang2019} divided the main burst of GRB 170114A into two time bins, and PA has $90^{\circ}$ change between the two time bins. For the same burst, it is divided into nine time bins, a roughly $90^{\circ}$ PA change happens between the second and third time bins, and a roughly $90^{\circ}$ PA change happens between the fifth and sixth time bins (Burgess et al. 2019). The PA in the single pulse burst GRB 170101A also have $90^{\circ}$ changes \citep{Wu2022}.

The physical quantities are parameterized and there is only one emitting region, the magnetic reconnection modle used in Uhm \& Zhang (2015), Uhm \& Zhang (2016), and Uhm et al. (2018) might be the simplest model for GRB prompt phase. Therefore, in this paper, we adopt this model for our polarization calculations. The paper is arranged as follows. In Section 2, we briefly introduce the polarization model, give our numerical results, and the understandings of the results. In Section 3, we apply our models to GRB 170101A and 170114A. Finally, we present our conclusions and discussion in Section 4.

\begin{figure*}[htb]
    \centering
    \includegraphics[width=1\textwidth]{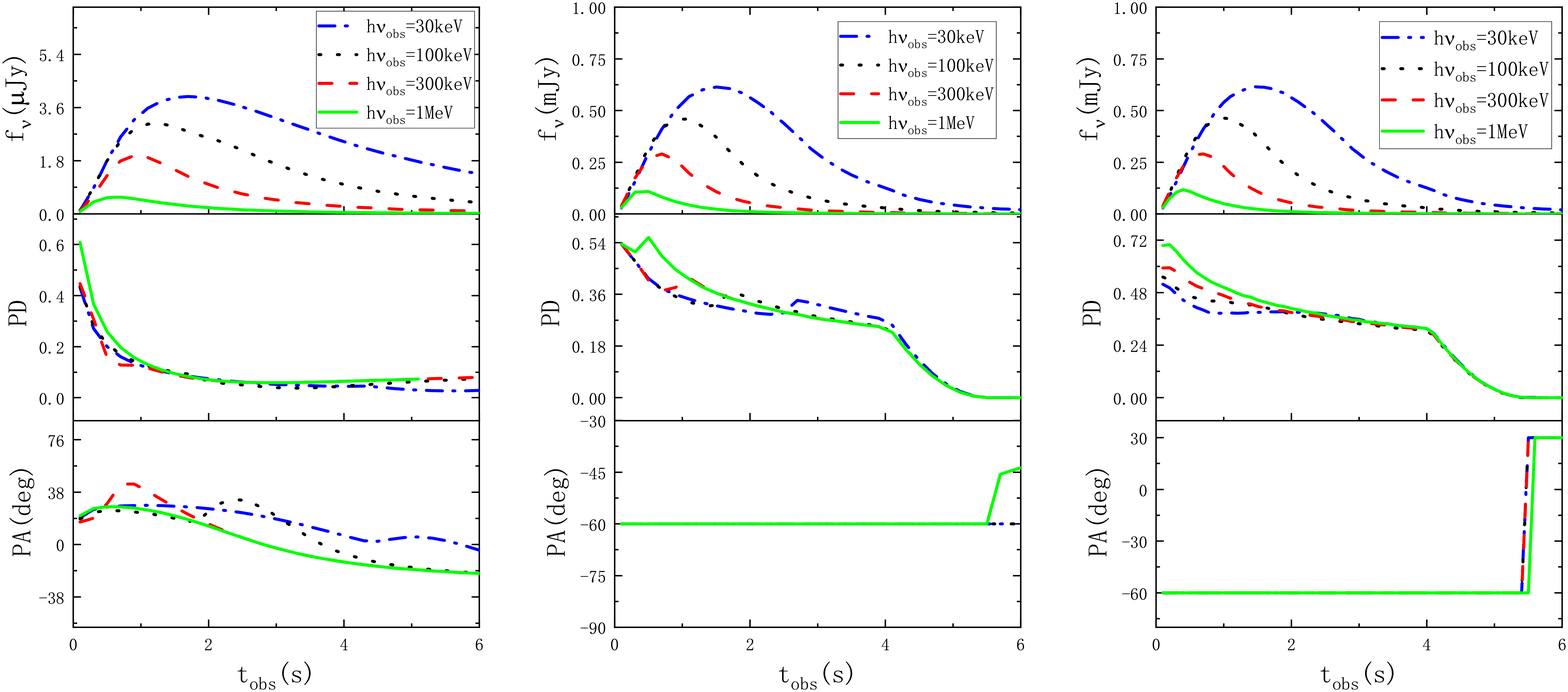}
    \caption{Light curves and polarization evolutions of $[2b_{i}]$ model. The top, middle, and bottom panels show the light curves, PD curves, and PA curves, respectively. The left and middle panels are calculated with the broken local PD, while the local PD of right panels is same as Lan \& Dai (2020). The observational angles for the left, middle, and right panels are 0.11 rad, 0 rad, and 0 rad, respectively. The green-solid, red-dashed, black-dotted, and blue-dashed-dotted lines correspond to the observational frequencies of 1 MeV , 300 keV, 100 keV, and 30 keV, respectively.
    }
    \label{tu1}
\end{figure*}

\begin{figure*}[htb]
    \centering
    \includegraphics[width=1\textwidth]{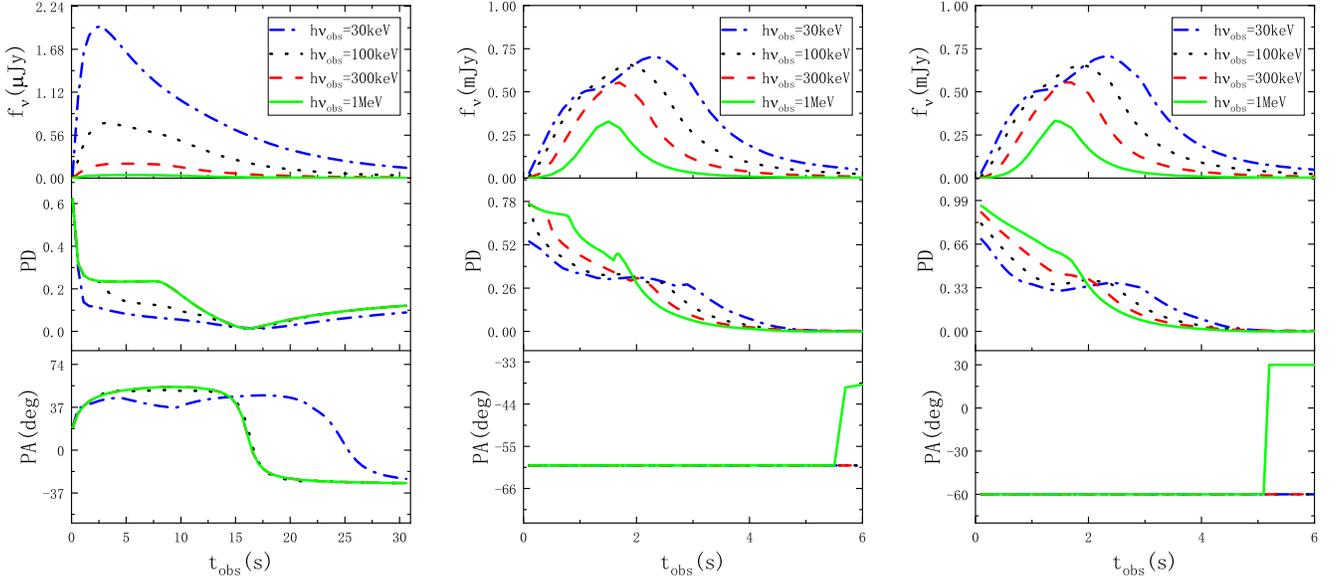}
    \caption{
        Same as Figure 1, but for model $[2b_{m}]$.
    }
    \label{tu2}
\end{figure*}

\begin{figure*}[htb]
    \centering
    \includegraphics[width=0.8\textwidth]{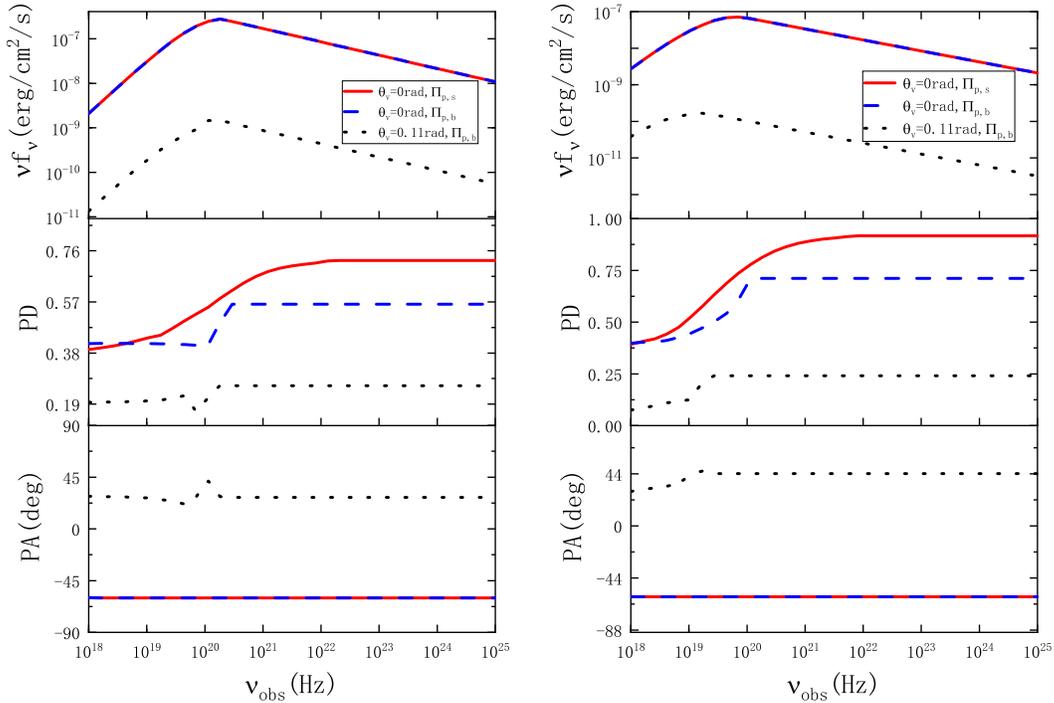}
    \caption{
        Polarization spectra for models $[2b_{i}]$ (left) and $[2b_{m}]$ (right). The top, middle, and bottom panels show the energy, PD, and PA spectra, respectively. For both left and right panels, the blue-dashed and black-dotted lines are calculated with the broken local PD, while local PD of red-solid lines is same as Lan \& Dai (2020). The observational angles of red-solid, blue-dashed, and black-dotted lines are 0 rad, 0 rad, and 0.11 rad, respectively. The observational times of red-solid, blue-dashed, and black-dotted lines for left panels are all 0.5 s. The observational times of red-solid, blue-dashed, and black-dotted lines for right panels are 0.5 s, 0.5 s, and 2.0 s, respectively.
    }
    \label{tu3}
\end{figure*}

\begin{figure*}[htb]
    \centering
    \includegraphics[width=0.8\textwidth]{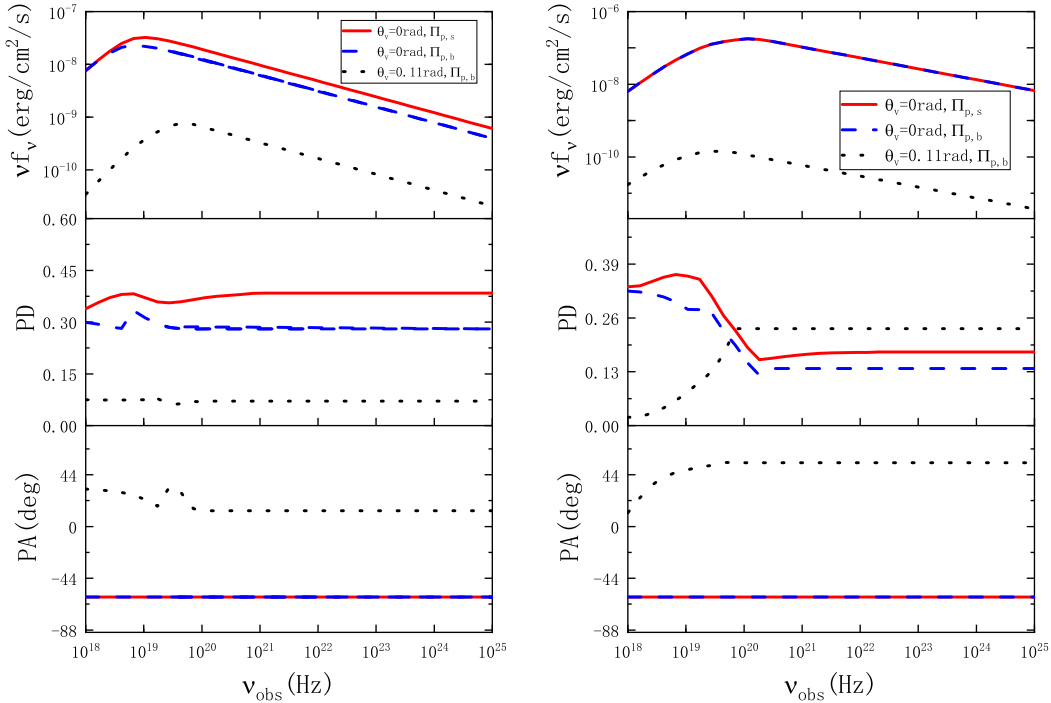}
    \caption{
        Same as Figure 3, but the observational times of red-solid, blue-dashed, and black-dotted lines in the left panel are 2.5 s, 2.9 s, and 2.0 s, respectively. The observational times of red-solid, blue-dashed, and black-dotted lines in the right panel are 2.5 s, 2.5 s, and 8.0 s, respectively.
    }
    \label{tu4}
\end{figure*}

\begin{figure*}[htb]
    \centering
    \includegraphics[width=0.6\textwidth]{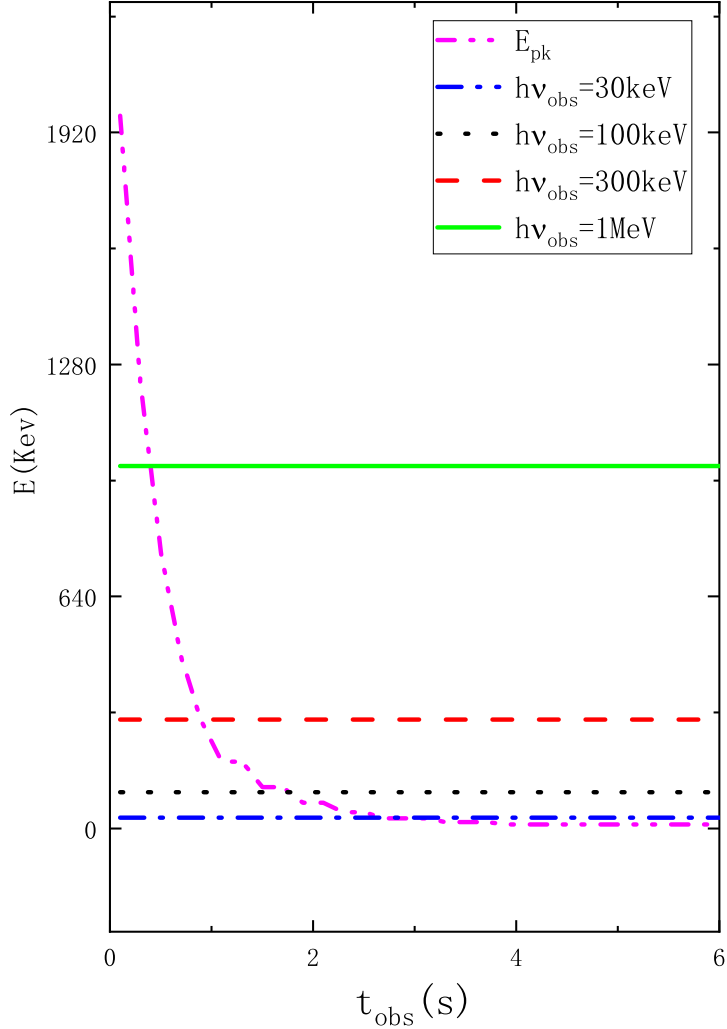}
    \caption{
        The peak energy evolutions of $[2b_{i}]$ model. The magenta-double-dotted-dashed, green-solid, red-dashed, black-dotted, and blue-dashed-dotted lines correspond to the peak energy, the observational frequencies of 1 MeV , 300 keV, 100 keV, and 30 keV, respectively.
    }
    \label{rho_B_ch5}
\end{figure*}

\begin{figure*}[htb]
    \centering
    \includegraphics[width=1\textwidth]{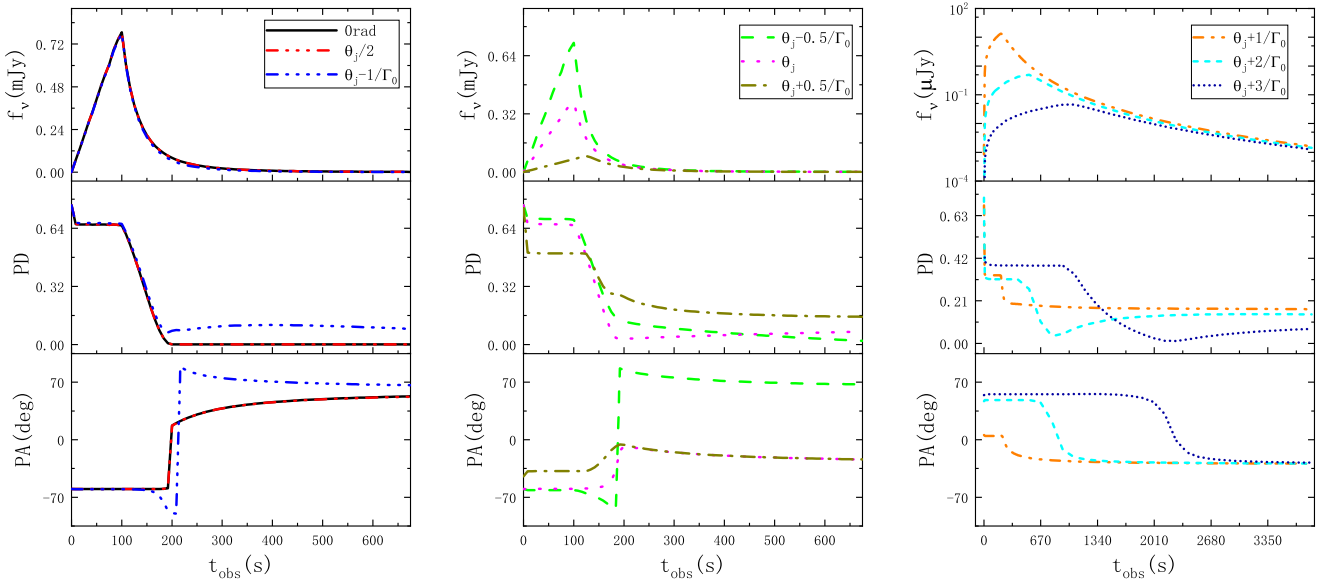}
    \caption{
        Light curves and polarization evolutions. The observational energies here is 300 keV. The top, middle, and bottom panels show the light curves, PD curves, and PA curves, respectively. We take $\Gamma_{0}=100$ and $s=b=g=0$. The black-solid, red-double-dashed, blue-three-dashed, green-dashed, magenta-dotted, dark-yellow-dashed-dotted, orange-double-dotted-dashed, cyan-short-dashed, and dark-blue-short-dotted lines correspond to $\theta_{V}=0$ rad, $\theta_{V}={\theta_{j}}/{2}$, $\theta_{V}=\theta_{j}-{1}/{\Gamma_{0}}$, $\theta_{V}=\theta_{j}-{0.5}/{\Gamma_{0}}$, $\theta_{V}=\theta_{j}$, $\theta_{V}=\theta_{j}+{0.5}/{\Gamma_{0}}$, $\theta_{V}=\theta_{j}+{1}/{\Gamma_{0}}$, $\theta_{V}=\theta_{j}+{2}/{\Gamma_{0}}$, and $\theta_{V}=\theta_{j}+{3}/{\Gamma_{0}}$, respectively.
    }
    \label{tu6}
\end{figure*}

\begin{figure*}[htb]
    \centering
    \includegraphics[width=1\textwidth]{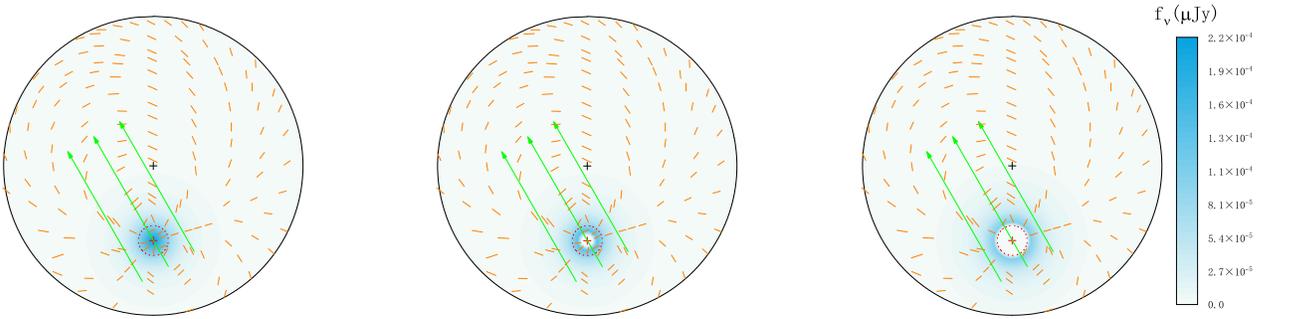}
    \caption{
        Schematics of flux and polarization on the jet sky plane. The observational energy is 300 keV and $\theta_{V}={\theta_{j}}/{2}$. The observational time of the left, middle and right schematics are 80 s, 120 s and 200 s, respectively. The edge of the jet cone is indicated by a black circle, and the black plus symbol denotes the jet symmetry axis. The observer's LOS is marked with a red plus symbol. The blue shadow shows the flux distribution. The red-dotted circle shows the edge of ${1}/{\Gamma_{0}}$ cone. The length of orange lines show the value of PD. The angle between the orange lines and the horizontal axis represent the value of PA. The direction of the green lines indicate the direction of the magnetic field.
    }
    \label{tu7}
\end{figure*}

\begin{figure*}[htb]
    \centering
    \includegraphics[width=1\textwidth]{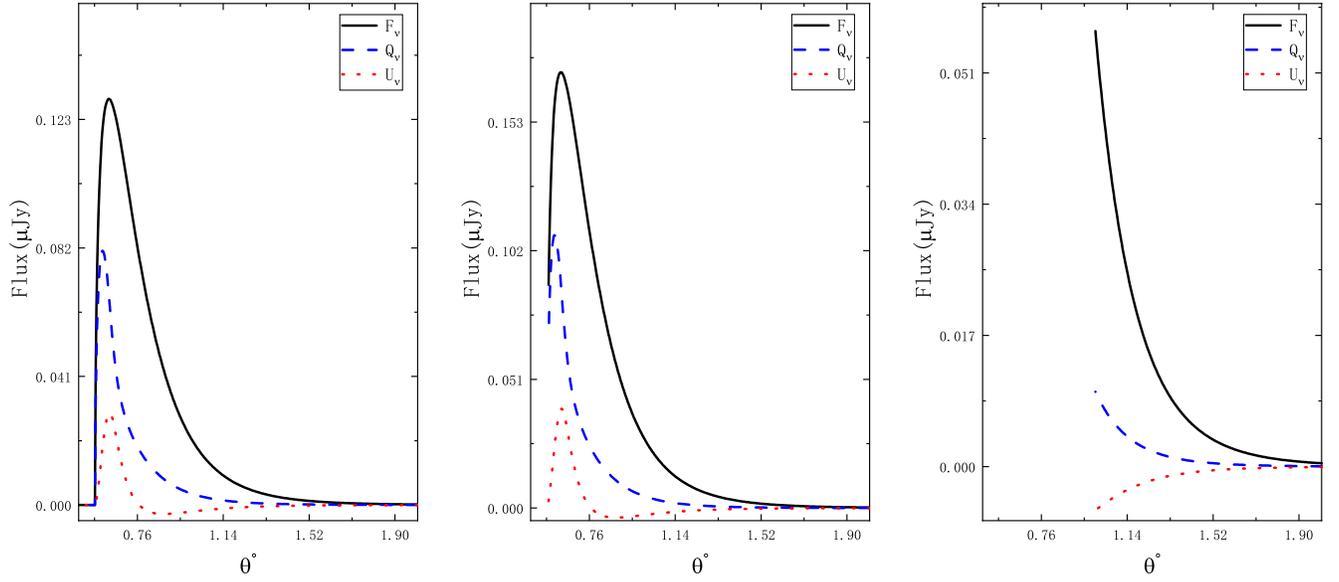}
    \caption{
        Stokes parameters evolution with $\theta$ for $\theta_{V}=\theta_{j}+{1}/{\Gamma_{0}}$ in Figure 6. The observational time of the left, middle and right schematics are 150 s, 200 s and 400 s, respectively. The black-solid, blue-dashed, and red-dotted lines correspond to the Stokes parameters $F_{\nu}$, $Q_{\nu}$ and $U_{\nu}$, respectively.
    }
    \label{tu8}
\end{figure*}

\begin{figure*}[htb]
    \centering
    \includegraphics[width=1\textwidth]{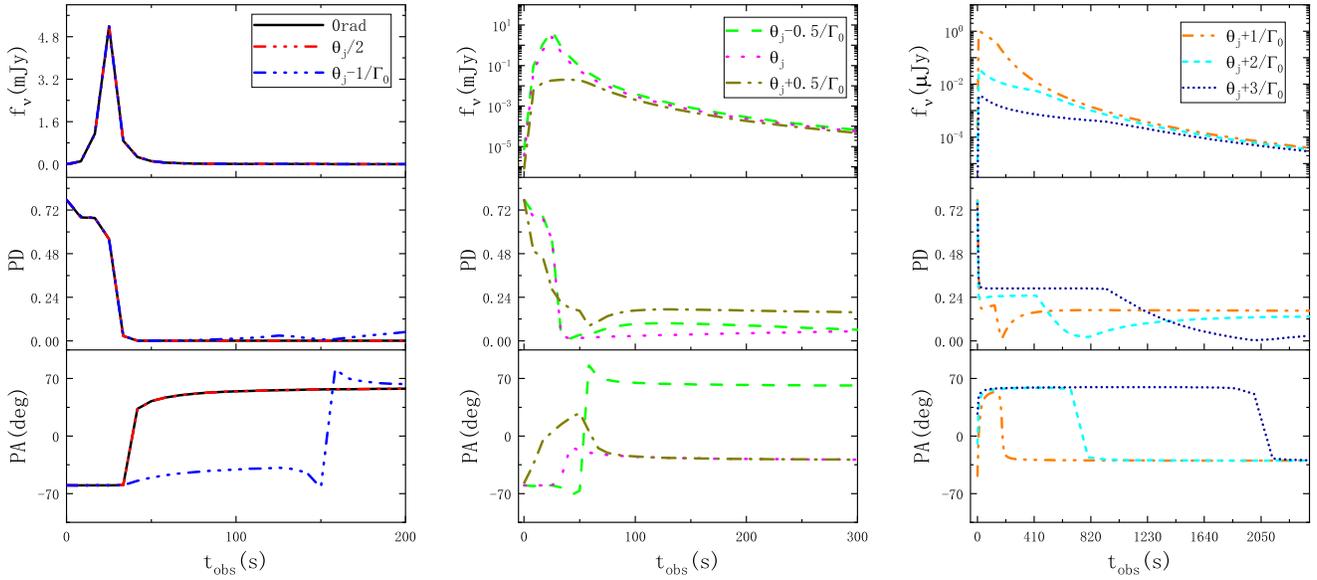}
    \caption{
        Same as Figure 6, but we take $s=0.35$.
    }
    \label{tu9}
\end{figure*}

\begin{figure*}[htb]
    \centering
    \includegraphics[width=0.8\textwidth]{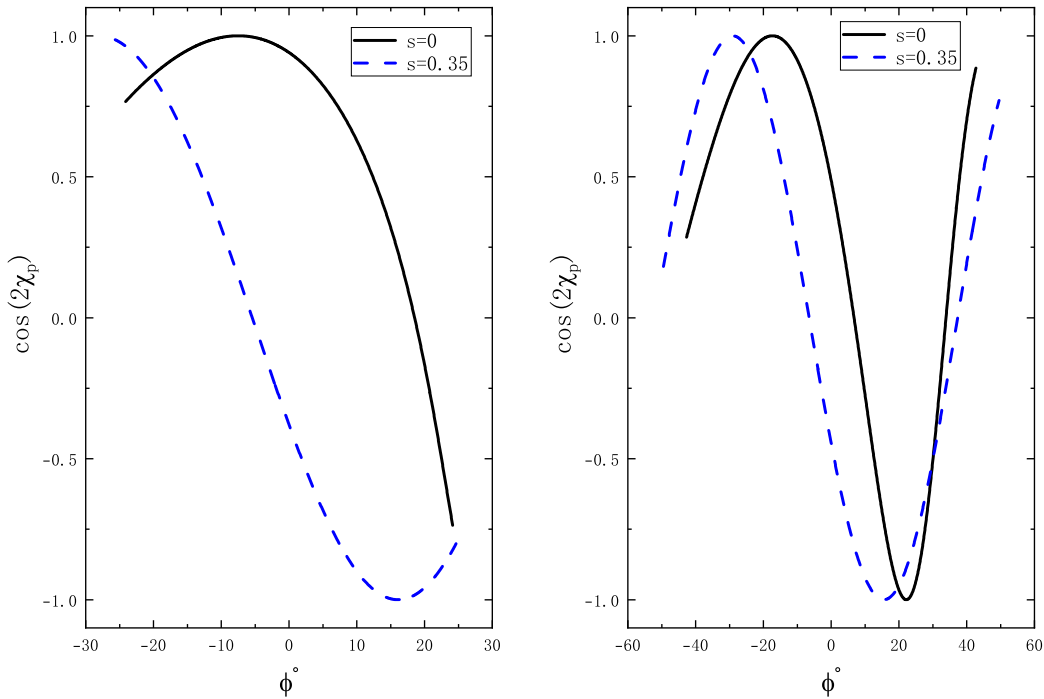}
    \caption{
        Evolution of $\cos(2\chi_{p})$ value with azimuth ($\phi$), the corresponding observational time of the left and right panels are 150 s and 300 s, respectively. The corresponding observational angle of the left and right panels are $\theta_{V}=\theta_{j}+1/{\Gamma_{0}}$. The corresponding $\theta_{p}$ of the black-solid and blue-dashed lines in the left panel are $0.63^{\circ}$ and $0.64^{\circ}$, respectively. The corresponding $\theta_{p}$ of the black-solid and blue-dashed lines in the right panel are $0.82^{\circ}$ and $0.95^{\circ}$, respectively. The black-solid and blue-dashed lines correspond to $s=0$ and $s=0.35$, respectively. 
    }
    \label{tu10}
\end{figure*}

\begin{figure*}[htb]
    \centering
    \includegraphics[width=1\textwidth]{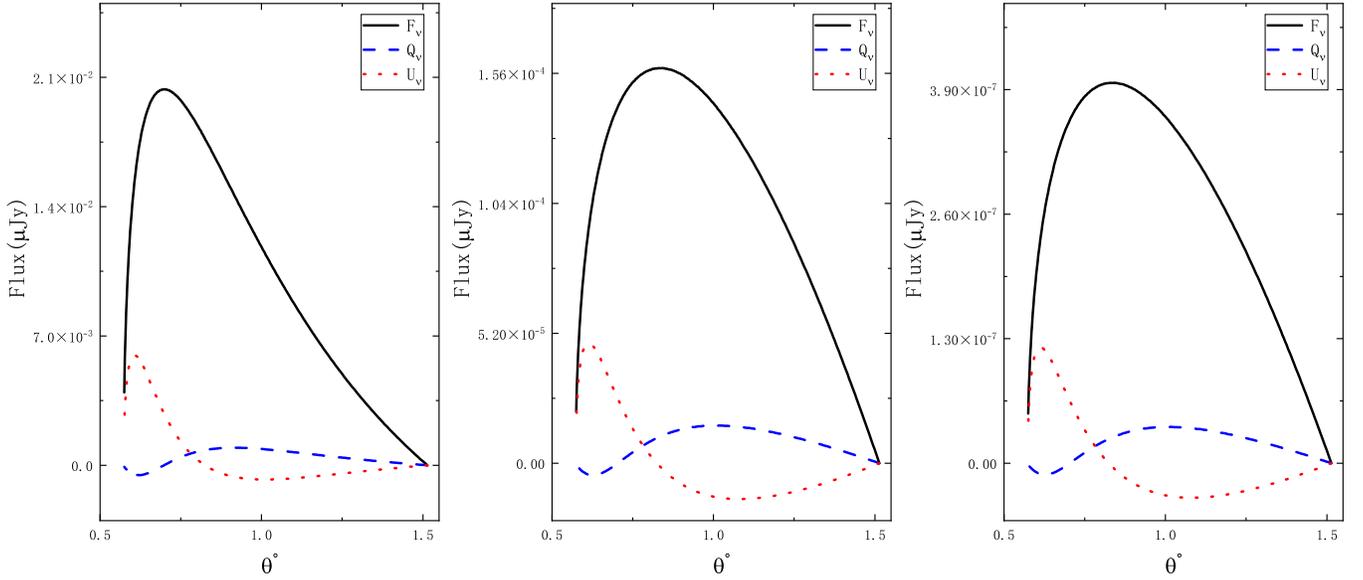}
    \caption{
        Stokes parameters evolution with $\theta$. The observational time is 2.0 s and $\theta_{V}=0.11$ rad. The corresponding observational frequencies of the left, middle, and right panels are $10^{18}$ Hz, $10^{21}$ Hz, and $10^{23}$ Hz, respectively. The black-solid, blue-dashed, and red-dotted lines correspond to the Stokes parameters $F_{\nu}$, $Q_{\nu}$ and $U_{\nu}$, respectively.
    }
    \label{tu11}
\end{figure*}

\begin{figure*}[htb]
    \centering
    \includegraphics[width=0.6\textwidth]{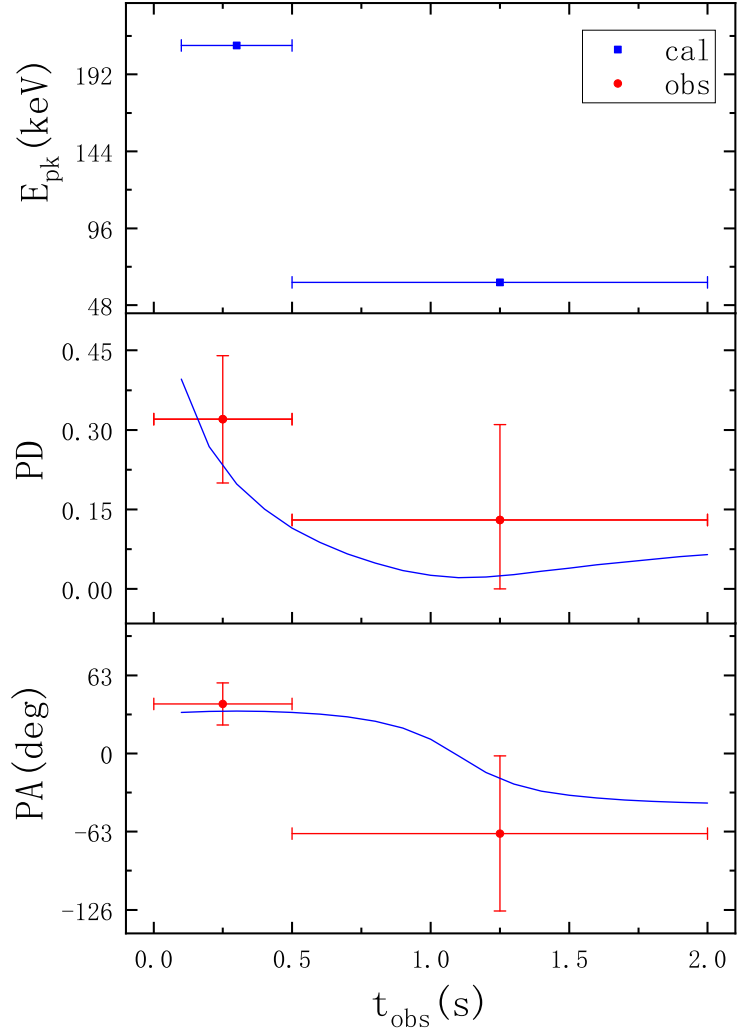}
    \caption{
        Evolutions of the peak energy (upper panel), PDs (middle panel), and PAs (lower panel) of GRB 170101A. The red circles and blue squares corresponds to the observational datas and  the theoretical calculation datas, respectively. The blue lines in middle and bottom panels represent our corresponding predicted curves.
    }
    \label{tu12}
\end{figure*}

\begin{figure*}[htb]
    \centering
    \includegraphics[width=0.6\textwidth]{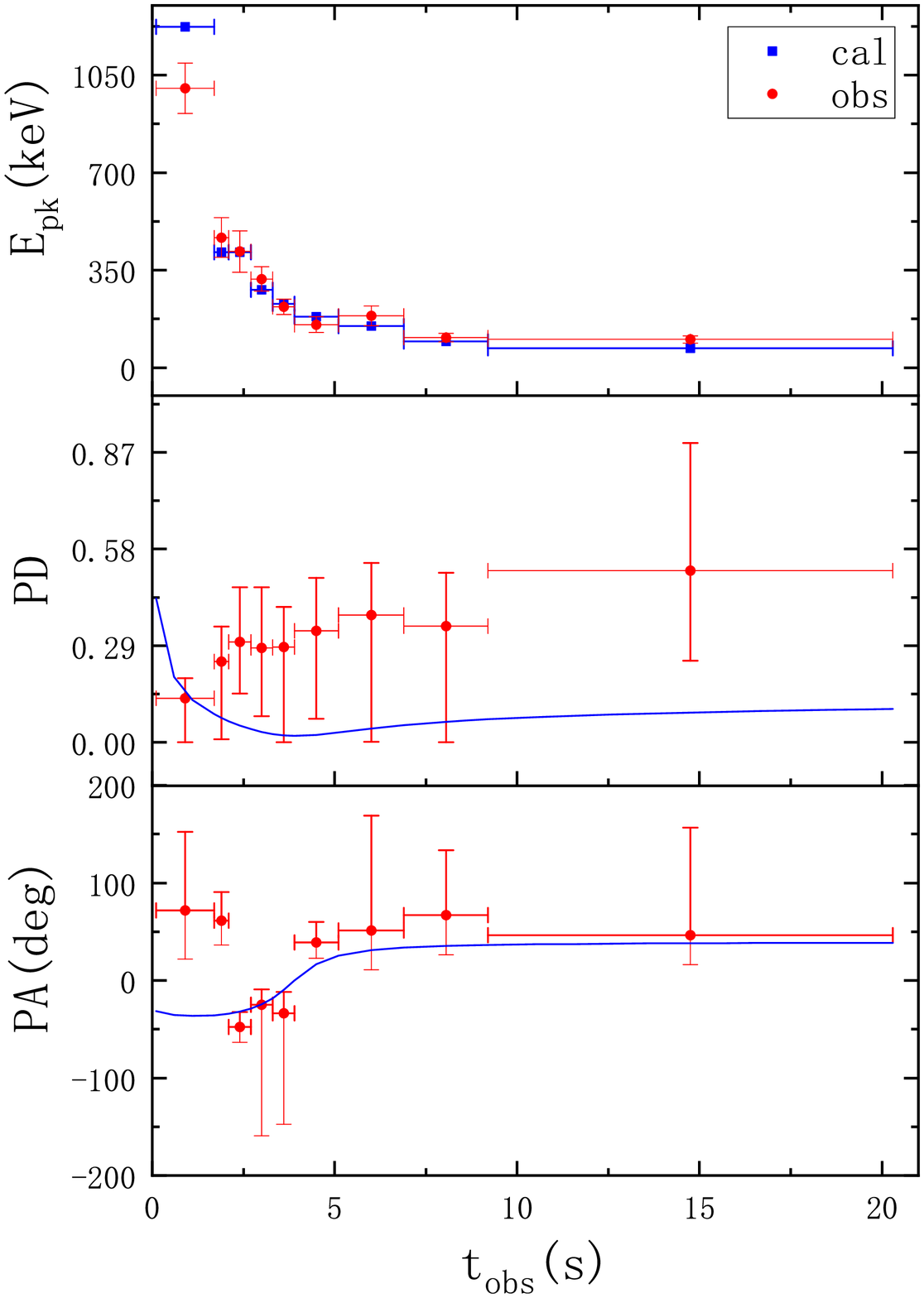}
    \caption{
        Same as Figure \ref{tu12}, but for GRB 170114A.
    }
    \label{tu13}
\end{figure*}

\section{The Model and the numerical results}\label{emodels}

\subsection{The model}\label{models}
We adopt a simple physical picture as Uhm \& Zhang (2015), Uhm \& Zhang (2016) and Uhm et al. (2018). A thin relativistic spherical shell expands radially in space and continuously emits photons from all positions in the shell. An isotropic angular distribution of the radiation power in the co-moving frame of the shell is assumed. The polarization model we use can be found in \cite{Lan2020}.

The magnetic field brought by the outflow from the central engine is large-scale ordered. The magnetic reconnection can create magnetic islands and turbulence, so the magnetic field configuration of magnetic reconnection model is likely to be a mixed magnetic field with an ordered part. Lan \& Dai (2020) pointed out that the polarization properties of a mixed magnetic field with an aligned ordered part are very similar to those of a purely aligned ordered magnetic field, only with a smaller polarization degree (PD) value. Here we assume that the magnetic field configuration is a large-scale aligned magnetic field, and the calculated PD gives the upper limit of the PD in a corresponding mixed magnetic field.

In the fluid co-moving frame, the critical frequencies in Equation (2) of Lan \& Dai (2020) were written incorrectly, and it should read as follows

\begin{equation}
    \quad \nu_{ch}^{\prime}=\frac{q_{e}B^{\prime}\gamma_{ch}^{2}}{2\pi m_{e}c}
\end{equation}
where $m_{e}$ and $q_{e}$ is mass and charge of the electron, respectively. $c$ represents the speed of light. $B^{\prime}$ represents the magnetic field strength in the co-moving system, and $\gamma_{ch}$ denotes the Lorentz factor of the electron.

It is assumed that the shell starts to emit photons at the radius $r_{on}$ (at the burst \textbf{source} time $t_{on}$). Therefore, for off-axis observation ($\theta_{V}\geq\theta_{j}$), the photons emitted at the radius $r$ will reach the observer at observer time $t_{obs}$

\begin{equation}
    t_{obs}=[t-\frac{r}{c}\cos\theta-t_{on}+\frac{r_{on}}{c}\cos(\theta_{V}-\theta_{j})](1+z).
\end{equation}
where $\theta_{V}$ is the observed angle, $\theta_{j}$ is the jet half angle, and $\theta$ is the angle between the velocity of the jet element and the line of sight (LOS) in the observer frame.

Lan \& Dai (2020) only considers the case of $\theta_{V}=0$ \textbf{in} Equation (2), here we can calculate the cases with off-axis observations. Another difference from Lan \& Dai (2020) is that local PD used here is

\begin{equation}
    \Pi_{p,b} = \left\{
      \begin{array}{ll}
        (-\alpha_{B})/(-\alpha_{B}+\frac{2}{3}), & x\leq\alpha_{B}-\beta_{B},\\
        (-\beta_{B})/(-\beta_{B}+\frac{2}{3}), & x\geq\alpha_{B}-\beta_{B},
      \end{array}
    \right.
  \end{equation}
we call this local PD as broken local PD ($\Pi_{p,b}$) \textbf{which is} different from the local PD for single-energy electrons ($\Pi_{p,s}$) in Lan \& Dai (2020)

\subsection{Numerical results}\label{results}
We consider two models in Uhm et al. (2018),ie., $[2b_{i}]$ and $[2b_{m}]$, and discuss their polarization properties. The only difference for the ``i'' and ``m'' models is their profiles of electron Lorentz factor $\gamma_{ch}$. The shell begins to emit photons at the radius $r_{on}$ and ceases at radius $r_{off}$. The bulk Lorentz factor is assumed to be a power-law form with radius, it can be expressed by (Drenkhahn 2002)

\begin{equation}
    \Gamma(r)=\Gamma_{0}(r/r_{0})^{s}.
\end{equation}

In the co-moving frame, the magnetic field strength of the radiated region decays with radius (Uhm \& Zhang 2014; Drenkhahn 2002),

\begin{equation}
    B^{\prime}(r)=B_{0}(r/r_{0})^{-b}.
\end{equation}

The properties of $\gamma_{ch}$ for ``i'' and ``m'' models can be described by the functions proposed by (Uhm \& Zhang 2014). It is assumed to be a single power law for ``i'' models,

\begin{equation}
    \gamma_{ch}(r)=\gamma_{ch}^{0}(r/r_{0})^{g},
\end{equation}
where we take $\gamma_{ch}^{0}=5\times10^{4}$, and $g=-0.2$. For ``m'' models, the electron Lorentz factor is assumed to be a broken power law and it reads

\begin{equation}
    \gamma_{ch}(r)= \gamma_{ch}^{m}\times\left\{
      \begin{array}{ll}
        (r/r_{m})^{g}, & r\leq r_{m},\\
        (r/r_{m})^{-g}, & r\geq r_{m},
      \end{array}
    \right.
  \end{equation}
and we take $\gamma_{ch}^{m}=2\times10^{5}$, $r_{m}=2\times10^{15}$ cm, and $g=1.0$.

The model parameters we take are same as Uhm et al. (2018), $\alpha_{B}=-0.8$, $\beta_{B}=-2.3$, $\Gamma_{0}=250$, $s=0.35$, $r_{on}=10^{14}$ cm, $r_{off}=3\times10^{16}$ cm, $r_{0}=1\times10^{15}$ cm and $B_{0}=30$ G. The injection rate of electrons in the shell is assumed to be $R_{inj}=10^{47}$ ${s}^{-1}$. The power-law index of the magnetic field strength is assumed to be $b=1.0$ for $[2b_{i}]$ and $[2b_{m}]$. The half-opening angle of the jet is taken as $\theta_{j}=0.1$ rad. The orientation of the aligned magnetic field is assumed to be $\delta=\pi/6$.

Figure \ref{tu1} shows our results for $[2b_{i}]$ models. In the first column, the observational angle is 0.11 rad and the local PD is $\Pi_{p,b}$. In the second column, the observational angle is 0 rad and the local PD is $\Pi_{p,b}$, and the observational angle is 0 rad and the local PD is $\Pi_{p,s}$ in the third column, same as Lan \& Dai (2020). The only difference between the first column and the second column is the observational angle. For off-axis observation shown in the first column, PDs decay fast with time and PAs evolve with time gradually for all calculated energy bands. The only difference between the second column and the third column is the local PD. For on-axis observations ($\theta_{V}\leq\theta_{j}$) with $\Pi_{p,s}$, PD decays with time, as shown in the third column, while for on-axis observations with $\Pi_{p,b}$, PDs decay in general with time, but show small bumps. For the observational frequencies of 30 keV, 100 keV, 300 keV, and 1 MeV, the observational times of the PD bump are 2.7 s, 1.7 s, 1.1 s, 0.5 s, respectively. The reason for these PD bumps will be interpreted in Section 2.3. For on-axis observations with both $\Pi_{p,b}$ and $\Pi_{p,s}$, PAs remain constants during the main burst.

Figure 2 is same as Figure 1, but for $[2b_{m}]$ model. The only difference between the first column and the second column in Figure 2 is the observational angle. For off-axis observation in the first column, PDs roughly decay with time and PAs evolve with time for all calculated energy bands. For on-axis observations with $\Pi_{p,s}$ in the third column, PD decays with time, while for on-axis observations with $\Pi_{p,b}$ in the second column, PDs roughly decay with time, but for the observational frequencies of 1 MeV, PDs show a sudden rise at $t_{obs}=1.7$ s. For on-axis observations with both $\Pi_{p,b}$ and $\Pi_{p,s}$, PAs remain constants during the main burst.

Figures \ref{tu3} and \ref{tu4} show the polarization spectra. Although different local PD is used here, for both models, the evolution trends of PDs are also increasing with energy range from soft X-ray to GeV $\gamma$-ray for different observational angles at the early observational time as in Lan \& Dai (2020). At late evolution times, for $[2b_{i}]$ model the evolution trends of PDs are not obvious. For the $[2b_{m}]$ models, PDs increase with energy for on-axis observations, while they decrease for off-axis observations. PAs are constants within the calculated energy band for on-axis observations, but show variations with energy for off-axis observations.

\subsection{Understanding of the results}\label{Understanding}

In Section 2.2, we have shown that PD will show a sudden rise at a certain time for on-axis observations with $\Pi_{p,b}$. To interpret this, we do some numerical calculations and the parameter settings are same as we used in Figure 1. We calculate the peak energy evolution of $[2b_{i}]$ model for on-axis observation and the results are shown in Figure 5.

In Figure 5, the peak energy is the peak of $\nu F_{\nu}$ spectrum. We find that $E_{pk}$ decays with time, and it intersect with the observational frequencies of 1 MeV, 300 keV, 100 keV, and 30 KeV at $t_{obs}=0.5$ s, $t_{obs}=1.1$ s, $t_{obs}=1.7$ s, and $t_{obs}=2.7$ s, respectively. For the observational frequencies of 1 MeV, 300 keV, 100 keV, and 30 KeV, PD bumps for $[2b_{i}]$ model are also around $t_{obs}=0.5$ s, $t_{obs}=1.1$ s, $t_{obs}=1.7$ s, and $t_{obs}=2.7$ s, respectively, as shown in the second column of Figure 1. This is mainly due to the dependence of the synchrotron polarization on the spectral indices (Equation 3). $\alpha_{B}$ and $\beta_{B}$ are the low-energy and high-energy spectral indices of the photon number flux, respectively. $\alpha_{B}$ is usually larger than $\beta_{B}$. So from Equation 3, we know the local PD $\Pi_{p,b}$ of low-energy photons is smaller than that of high-energy photons. When $E_{pk}$ crosses the observational frequency at some observational time,
high-energy photons with larger local PD will contribute more to radiation. So the PD becomes larger around the crossing observational time than the adjacent observational times.

In Section 2.2, PAs will evolve with time for off-axis observations. We perform some numerical calculations to interpret these PA rotations. In the following, same as in Section 2.2, we take $\alpha_{B}=-0.8$, $\beta_{B}=-2.3$, $R_{inj}=10^{47}$ ${s}^{-1}$, $r_{on}=10^{14}$ cm, $r_{off}=3\times10^{16}$ cm, $r_{0}=1\times10^{15}$ cm, $B_{0}=30$ G, $\gamma_{ch}^{0}=5\times10^{4}$, $\theta_{j}=0.1$ rad, and $\delta=\pi/6$. Here, we calculate the light curves and polarization evolutions for $\Gamma_{0}=100$, 250, 800, respectively. And we find the profiles of both the light curves and polarization curves for different $\Gamma_{0}$ are similar. So for illustration, we take $\Gamma_{0}=100$ (small contrast between $1/\Gamma_{0}$ cone and the jet cone) as example to interpret our results in Section 2.2.

First, we set $s=b=g=0$ to ensure that the magnetic field strength, bulk Lorentz factor and electron Lorentz factor are all invariants on one equal arrival time surface (EATS). With the above parameters, we calculate light curves and polarization evolutions shown in Figure 6. We find that PAs remain constants for $\theta_{V}\geq\theta_{j}+2/{\Gamma_{0}}$, while it evolves with time for $\theta_{V}<\theta_{j}+2/{\Gamma_{0}}$. For illustration, taking $\theta_{V}=\theta_{j}/2$ as an example, its PA curve show an abrupt change around $t_{obs}=200$ s. So we draw the schematics of flux and polarization on the jet sky plane for $t_{obs}=80$ s, $t_{obs}=120$ s and $t_{obs}=200$ s, respectively. The results are shown in Figure 7. The fluxes at the first two observational times are dominated by low-latitude radiation within ${1/\Gamma}$ cone, while it is dominated by high-latitude emission at third observational time.

On one EATS, with the increase of the radius $r$, $\theta$ will decrease. $\tilde{f}(t_{obs})$ is the ratio of the flux from $\theta$ circles between $\theta\Gamma(r)<1$ and $\theta\Gamma(r)>1$, it can be expressed by (Lan \& Dai 2020)

\begin{equation}
    \tilde{f}(t_{obs})=\frac{\int_{r_{c}}^{r_{max}}df_{\nu}}{\int_{r_{min}}^{r_{c}}df_{\nu}}.
\end{equation}
where $r_{max}=r(t_{obs},\theta=\theta_{min})$, $r_{min}=r(t_{obs},\theta=\theta_{max})$ and $\theta\Gamma(r_{c})=1$. $r_{max}$ and $r_{min}$ are the maximum and minimum radius on the EATS with observational time $t_{obs}$, respectively. On one EATS (corresponding to an observational time), $\tilde{f}(t_{obs})>1$ means that low-latitude emission dominates the jet radiation, while $\tilde{f}(t_{obs})<1$ means that high-latitude emission dominate the jet radiation. With the calculation, $\tilde{f}(t_{obs})>1$ for $t_{obs}=80$ s and $t_{obs}=120$ s, while it is 0 for 200 s in Figure 7.

The values of $\tilde{f}(t_{obs})$ at $t_{obs}=150$ s, $t_{obs}=200$ s and $t_{obs}=400$ s for $\theta_{V}=\theta_{j}+1/{\Gamma_{0}}$ in Figure 6 are all zero. This means that the radiations from these three EATSs are dominated by high-latitude emission. However, its PA curve still show gradual rotation after 200 s. Then we calculate the variations of Stokes parameters with $\theta$, as shown in Figure 8. We find the Stokes parameters have a sharp peak on one EATS, the corresponding $\theta$ is denoted as $\theta_{p}$. So the radiation from $\theta_{p}$ circle dominate the radiation on this EATS. $\theta_{p}$s for $t_{obs}=150$ s and 200 s are the same and equal to $\theta_{p}=0.64^{\circ}$, while it changes to $0.99^{\circ}$ for $t_{obs}=400$ s. As mentioned above, PA curve of $\theta_{V}=\theta_{j}+1/{\Gamma_{0}}$ stays as constant before 200 s and shows gradual rotation after roughly 200 s. So when the value of $\tilde{f}(t_{obs})$ remains zero, if the value of $\theta_{p}$ changes, PA will rotate, and if $\theta_{p}$ is unchanged, PA will stay as constant.

Secondly, we set $b=g=0$ and $s=0.35$. With other fixed parameters mentioned above, we calculate the light curves and polarization evolutions for various observational angle as shown in Figure 9. The only difference of Figure 9 from Figure 6 is that $\Gamma(r)$ evolves with radius. We find that PAs evolve with time for all the observational angles calculated. PAs show abrupt changes after main bursts for on-axis observations, and they evolve more violently for off-axis observations than that for $s=0$ in Figure 6.

On one EATS, the emission from $\theta_{p}$ circle dominate the radiation on this EATS. With the calculation, we find $\theta_{p}=0.63^{\circ}$ at $t_{obs}=150$ s, and $\theta_{p}=0.82^{\circ}$ at $t_{obs}=300$ s for $\theta_{V}=\theta_{j}+{1}/{\Gamma_{0}}$ with $s=0$ in Figure 6. $\theta_{p}=0.64^{\circ}$ at $t_{obs}=150$ s, and $\theta_{p}=0.95^{\circ}$ at $t_{obs}=300$ s for $\theta_{V}=\theta_{j}+{1}/{\Gamma_{0}}$ with $s=0.35$ in Figure 9. Since the local Stokes parameter $U_{\nu}(\theta)\propto\sin(2\chi_{p})$ and $Q_{\nu}(\theta)\propto\cos(2\chi_{p})$, to interpret the more violent PA changes in $s=0.35$, we take $s=0$ and $s=0.35$ respectively, and other fixed parameters mentioned above, we calculate the variations of $\cos(2\chi_{p})$ value with azimuth $\phi$ at different $\theta_{p}$ for $\theta_{V}=\theta_{j}+{1}/{\Gamma_{0}}$. The results are shown in Figure 10, we find an evolving bulk Lorentz factor $\Gamma(r)$ will lead to an increase of $\chi_{p}$ value. After integration over $\phi$, $Q_{\nu}(\theta_{p})$ might be negative. For $\theta_{V}=\theta_{j}+{1}/{\Gamma_{0}}$ with $s=0$ in Figure 6, the values of $Q_{\nu}(\theta_{p})$ at $t_{obs}=150$ s and $t_{obs}=300$ s are all positive. The values of $U_{\nu}(\theta_{p})$ at $t_{obs}=150$ s and $t_{obs}=300$ s are positive and negative, respectively. For $\theta_{V}=\theta_{j}+{1}/{\Gamma_{0}}$ with $s=0.35$ in Figure 9, the values of $U_{\nu}(\theta_{p})$ at $t_{obs}=150$ s and $t_{obs}=300$ s are also positive and negative, respectively. The value of $Q_{\nu}(\theta_{p})$ at $t_{obs}=300$ s is positive, while value of $Q_{\nu}(\theta_{p})$ at $t_{obs}=150$ s changes to positive. When $Q_{\nu}(\theta_{p})<0$, the final value of PA should be added ($U_{\nu}(\theta_{p})>0$) by $90^{\circ}$ (Lan et al. 2018). Therefor, the changing bulk Lorentz factor leads to PAs evolve more violently for off-axis observations.

We also study the influences of both the magnetic field strength $B^{\prime}$ and the electron Lorentz factor $\gamma_{ch}$ on the rotations of PAs. Other parameters are same as we used in Figure 6, we take $b=1$ and $g=-0.2$ to study, respectively. We find that the changes of $B^{\prime}$ and $\gamma_{ch}(r)$ have slight effects on the PA evolutions.

Figures 3 and 4 have shown that the PAs are also constants within the calculated energy band for on-axis observations, but it evolves for off-axis observations. For illustration, taking $[2b_{i}]$ model for $\theta_{V}=0.11$ rad with $t_{obs}=2.0$ s as an example, and parameters are same as we used in Figure 4. With the calculation, the values of $\tilde{f}(t_{obs})$ for $\nu_{obs}=10^{18}$ Hz, $\nu_{obs}=10^{21}$ Hz and $\nu_{obs}=10^{23}$ Hz in Figure 4 are all zero. However, PA still evolve within the calculated energy band. Then we calculate the variations of Stokes parameters with $\theta$, as shown in Figure 11. We find $\theta_{p}$s for $\nu_{obs}=10^{18}$ Hz and $\nu_{obs}=10^{21}$ Hz are different, and PAs for $\nu_{obs}=10^{18}$ Hz and $\nu_{obs}=10^{21}$ Hz are also different. The value of $\theta_{p}$ is $0.69^{\circ}$ for $\nu_{obs}=10^{18}$ Hz, and the vaule of $\theta_{p}$ changes to $0.84^{\circ}$ for $\nu_{obs}=10^{21}$ Hz. The value of PA is $31.9^{\circ}$ for $\nu_{obs}=10^{18}$ Hz, and the vaule of PA also change to $13.3^{\circ}$ for $\nu_{obs}=10^{21}$ Hz. While $\theta_{p}$s for $\nu_{obs}=10^{21}$ Hz and $10^{23}$ Hz are the same and equal to $\theta_{p}=0.84^{\circ}$, and PAs for $\nu_{obs}=10^{21}$ Hz and $10^{23}$ Hz are also the same and equal to $13.3^{\circ}$. So when the value of $\tilde{f}(t_{obs})$ remains zero, if the value of $\theta_{p}$ changes, PA will rotate, and if $\theta_{p}$ is unchanged, PA will stay as constant.

\section{Application to GRB 170101A and GRB 170114A}\label{application}

\subsection{GRB 170101A}\label{GRB1}

The observational data of this burst are taken from Kole et al. (2020). The main burst of GRB 170101A are divided into two time bins, the first and second time bins are 0.0-0.5 s and 0.5-2.0 s, respectively. We use the $[2b_{i}]$ model with the large-scale ordered aligned magnetic field configuration to fit GRB170101A. The parameters adopted in our fittings are $\theta_{j}=0.1$ rad, $\theta_{V}=0.11$ rad, $z=1$, $\alpha_{B}=-1.44$, $\beta_{B}=-2.49$, $R_{inj}=10^{47}$ ${s}^{-1}$, $\Gamma_{0}=250$, $s=0.35$, $r_{on}=4\times10^{13}$ cm, $r_{off}=8\times10^{14}$ cm, $B_{0}=30$ G, $b=1.0$, $\gamma_{ch}^{0}=1.8\times10^{5}$, $r_{0}=1\times10^{14}$ cm, $g=-0.2$, $\delta=\pi/4$ and the local PD used is $\Pi_{p,b}$. The time-integrated flux can be derived from the time-resolved flux, and its formula is

\begin{equation}
    F_{t-integrated}=\frac{\int_{t_{1}}^{t_{2}}F_{\nu}dt_{obs}}{\int_{t_{1}}^{t_{2}}dt_{obs}},
\end{equation}
where $F_{\nu}$ can be found in Equation (5) of Lan \& Dai (2020). $t_{1}$ and $t_{2}$ represent the minimum and maximum of each time bins, respectively. For example, $t_{1}=0.5$ s and $t_{2}=2.0$ s for the second time bin. Since the theoretical calculations have no value at $t_{obs}=0$ s, we start the calculation at $t_{obs}=0.1$ s, so $t_{1}=0.1$ and $t_{2}=0.5$ for the first time bin. Through equation (8) and the above parameters, we calculated the $E_{pk}$ of the two time bins and the results are shown in Figure 12 (upper panel). The burst has no reported time-resolved observational data of peak energy. The predicted peak energy evolution pattern is hard-to-soft.

The detection energy range of POLAR is 50-500 KeV. We use the energy-integrated Stokes parameters to calculate PDs and PAs, and the formulas are

\begin{equation}
    F_{\nu-integrated}=\frac{\int_{\nu_{1}}^{\nu_{2}}F_{\nu}d\nu}{\int_{\nu_{1}}^{\nu_{2}}d\nu},
\end{equation}

\begin{equation}
    Q_{\nu-integrated}=\frac{\int_{\nu_{1}}^{\nu_{2}}Q_{\nu}d\nu}{\int_{\nu_{1}}^{\nu_{2}}d\nu},
\end{equation}

\begin{equation}
    U_{\nu-integrated}=\frac{\int_{\nu_{1}}^{\nu_{2}}U_{\nu}d\nu}{\int_{\nu_{1}}^{\nu_{2}}d\nu},
\end{equation}
where $\nu_{1}=50$ KeV and $\nu_{2}=500$ KeV. $Q_{\nu}$ and $U_{\nu}$ can be found in Equations (6) and (7) of Lan \& Dai (2020). Through above formulas, we can calculate the energy-integrated PD and PA and the results are shown in Figure 10 (middle panel and lower panel, respectively). We find the theoretically calculated PD decays with time and it can fit the observed PD.

We extract PA values with the 1$\sigma$ credibility from Figure 6 in Kole et al. (2020) for each time bins. Since the polarization direction is invariant with n$\times180^{\circ}$ variations in PA (n is an integer), here we set the observed PAs to be in the range [-$90^{\circ}$, $90^{\circ}$] by adding or subtracting n$\times180$ from the values given by Kole et al. (2020). A roughly -$90^{\circ}$ PA change happens between the first and second time bins. Our model can interpret this violent PA variation.

\subsection{170114A}\label{GRB2}

The observational data of this burst are taken from Burgess et al. (2019). Burgess et al. (2019) divided the main burst into nine time bins, and the initial time of the burst was -0.2 s, which we set to be 0.1s. Thus the nine time bins read [0.1-1.7 s], [1.7-2.1 s], [2.1-2.7 s], [2.7-3.3 s], [3.3-3.9 s], [3.9-5.1 s], [5.1-6.9 s], [6.9-9.2 s], and [9.2-20.3 s]. We extract peak energy from $\nu F_{\nu}$ spectra (Figure 10 of Burgess et al. (2019)) for each time bins. We use the $[2b_{i}]$ model with the large-scale ordered aligned magnetic field configuration to fit the data of GRB170114A. The parameters adopted in our fittings are $\theta_{j}=0.1$ rad, $\theta_{V}=0.11$ rad, $z=1$, $\alpha_{B}=-0.83$, $\beta_{B}=-2.04$, $R_{inj}=10^{47}$ ${s}^{-1}$, $\Gamma_{0}=800$, $s=0.35$, $r_{on}=1\times10^{14}$ cm, $r_{off}=3\times10^{17}$ cm, $B_{0}=30$ G, $b=1.0$, $\gamma_{ch}^{0}=1.6\times10^{4}$, $r_{0}=1\times10^{16}$ cm, $g=-0.2$, $\delta=\pi-\pi/4.5$ and the local PD used is $\Pi_{p,b}$.

With Equation (8) and the above parameters, we calculated the $E_{pk}$ of these nine time bins and our results are shown in Figure 13 (upper panel). The predicted peak energy evolution pattern is hard-to-soft, which fits the observed peak energy evolution pattern. Except for the first and ninth time bins, the theoretically calculated $E_{pk}$ fit the observed $E_{pk}$ of other time bins.

We take $\nu_{1}=50$ KeV and $\nu_{2}=500$ KeV to calculate the energy-integrated PDs and PAs. The results are shown in Figure 10 (middle panel and lower panel, respectively). We find that PD decays with time. The theoretically calculated PD could roughly fit the observed PD curve.

Here we set the observed PAs to be in the range [-$90^{\circ}$, $90^{\circ}$] by adding or subtracting n$\times180^{\circ}$ from the values given by Burgess et al. (2019). A roughly -$90^{\circ}$ PA change happens between the second and third time bins, and a roughly $90^{\circ}$ PA change happens between the fifth and sixth time bins. Our model can interpret the violent PA variation between the fifth and sixth time bins, but it cannot account for the violent PA variation between the second and third time bins. Further detailed models should be considered.

\section{Conclusions and discussion}\label{conclusions}

We consider the magnetic reconnection model to investigate the rotations of PA in GRB prompt emission. For the large-scale ordered aligned magnetic field configuration, we find that PAs evolve with time (energy) for off-axis observations.

For the models in this paper, PDs roughly decay with time, but at a certain observational time, PDs will show a small bump for on-axis observations if power-law local PD $\Pi_{p,b}$ is used. This is mainly due to the dependence of the synchrotron polarization on the spectral indices. When the decaying peak energy crosses the observational frequency at some observational time, high-energy photons with larger local PD will contribute more to radiation. So the PD will increase around this crossing observational time.

PAs will evolve with time (energy) for off-axis observations. Rotations of the PAs with time (energy) for off-axis observations are the results of the changes of $\tilde{f}$ or $\theta_{p}$. PAs with $\tilde{f}(t_{obs})>0$ will stay roughly as a constant. PAs with $\tilde{f}(t_{obs})=0$ will be different from that with $\tilde{f}(t_{obs})>0$. When $\tilde{f}=0$, if $\theta_{p}$ vaule is unchanged with time, so does the PAs. If $\theta_{p}$ vaule is changed, PAs will rotate. Both $\tilde{f}$ and $\theta_{p}$ are related to the ``observed shape'' of the emitting region (before average). Therefore, we conclude that the change of the``observed shape'' of the emitting region (before average) will lead to PA rotation Moreover, the evolving bulk Lorentz factor ($\Gamma(r)$ $\propto$ $r^{0.35}$) will make PA evolutions more violently.

In this paper, only PA evolutions of the burst with single pulse are considered. We use the $[2b_{i}]$ model with the large-scale ordered aligned magnetic field configuration to fit the data of GRB170101A. We find the theoretically calculated PD decays with time and it can fit the observed PDs of GRB 170101A. The model can also interpret the violent PA variation between the first and second time bins of the burst. In addition, we also use $[2b_{i}]$ model to interpret the observations of GRB 170114A. The predicted peak energy evolution pattern is hard-to-soft, which could roughly fit the observed peak energy evolution pattern of GRB 170114A. The model can interpret the violent PA variation between the fifth and sixth time bins of GRB 170114A, but it cannot account for the violent PA variation between the second and third time bins of this burst, so further more detailed studies are needed.

\begin{acknowledgements}

{This paper is dedicated to the 70th anniversary of the physics of Jilin University. This work is supported by the National Natural Science Foundation of China (grant No. 11903014).}

\end{acknowledgements}

\end{document}